\documentclass[aip,pop,reprint,graphicx]{revtex4-1}
\usepackage{graphicx}
\usepackage[dvipdfm,colorlinks,linkcolor=blue,citecolor=blue,urlcolor=blue]{hyperref}
\draft 

\begin{document}


\title{Hot-electron refluxing enhanced relativistic transparency of overdense plasmas} 



\author{Yong Yu}
\affiliation{National Key Laboratory of Shock Wave and Detonation
Physics, Institute of Fluid Physics, China Academy of Engineering Physics,
Mianyang, 621900, People's Republic of China}

\author{Xiao-Ya Li}
\email[Electronic mail: ]{xylapril@gmail.com}
\affiliation{National
Key Laboratory of Shock Wave and Detonation Physics, Institute of
Fluid Physics, China Academy of Engineering Physics, Mianyang, 621900, People's Republic of China}

\author{Zi-Yu Chen}
\affiliation{National
Key Laboratory of Shock Wave and Detonation Physics, Institute of
Fluid Physics, China Academy of Engineering Physics, Mianyang, 621900, People's
Republic of China}

\author{Jia-Xiang Wang}
\affiliation{State Key Laboratory of Precision Spectroscopy, East
China Normal University, Shanghai 200062, People's Republic of China}

\author{Wen-Jun Zhu}
\email[Electronic mail: ]{wjzhu@caep.ac.cn}
\affiliation{National Key Laboratory of Shock Wave and Detonation
Physics, Institute of Fluid Physics, China Academy of Engineering Physics,
Mianyang, 621900, People's Republic of China}



\date{\today}

\begin{abstract}
A new phenomenon of enhancing the relativistic transparency of overdense plasmas by the influence of hot-electron refluxing has been found via particle-in-cell simulations. When a p-polarized laser pulse, with intensity below the self-induced-transparency(SIT) threshold, obliquely irradiates a thin overdense plasma, the initially opaque plasma would become transparent after a time interval which linearly relies on the thickness of the plasma. This phenomenon can be interpreted by the influence of hot-electron refluxing. As the laser intensity is higher than the SIT threshold, the penetration velocity of the laser in the plasma is enhanced when the refluxing is presented. Simulation data with ion motion considered is also consistent with the assumption that hot-electron refluxing enhances transparency. These results have potential applications in laser shaping.
\end{abstract}


\maketitle 

\section{Introduction}

Propagation of intense lasers in an overdense plasma  have attracted a lot of attention due to its potential applications in the fast ignitor concept for inertial confinement\cite{tabak:1626}, charged-particles acceleration~\cite{PhysRevLett.103.045002}, laser shaping~\cite{vshivkov:2727}, neutron acceleration~\cite{PhysRevLett.110.044802} and so on. From the propagation condition $\omega>\omega_p$ , where  $\omega$ is the laser frequency and $\omega_p=(4\pi n_e e^2/m_e)^{1/2}$ is the plasma frequency ($n_e$ is the plasma density, $e$ and $m$ are the electron charge and mass, respectively), one can see that two approaches can be used to lower the effective plasma frequency in order to make the overdense plasma transparent for the laser light. One approach is to reduce the plasma density $n_e$ by means such as hole boring~\cite{PhysRevLett.69.1383,PhysRevLett.73.664}. And another is to increase the relativistic factor $\gamma$ of the electrons as well as the effective mass $m_e=\gamma m_0$ of the electrons, such as self induced transparency (SIT)\cite{kaw:472}. These studies mainly concern the effects of laser-plasma interaction in the front surface of the target. In this paper, it will be shown that the  hot-electron refluxing would participate in the surface laser-plasma interaction and have an effect on the transparency property of overdense plasmas.

When an ultraintense laser pulse irradiates a thin overdense plasma, a large number of hot electrons with velocity near the light speed can be generated. These hot electrons can propagate through the plasma slab and then return to the front surface of the plasma reflected by the strong sheath field on the rear surface of the thin plasma. These returned hot electrons are known as hot-electron refluxing, or recirculation~\cite{PhysRevLett.88.215006, sentoku:2009}. The hot-electron refluxing may play an important role in  proton acceleration~\cite{PhysRevLett.88.215006} and $K\alpha$ generation~\cite{neumayer:103103}. Recent experiments have also showed the importance of the lateral electron recirculation in small targets~\cite{PhysRevLett.105.015005}. However, the influence of the hot-electron refluxing on SIT was rarely considered in the previous studies.

In our work, we report a new phenomenon that hot-electron refluxing can enhance the self induced transparency of overdense plasmas. Using particle-in-cell (PIC) simulations, the transition from opacity to transparency in the overdense plasma is shown. The generation and propagation of the hot-electron refluxing are carefully investigated, which give a clue of how refluxing enhances transparency. Simulation results considering higher-intensity laser irradiation  and ion motion also verify the conclusion that the hot-electron refluxing enhances transparency. The results are potentially useful for applications such as laser shaping.

\section{TRANSITION FROM OPACITY TO TRANSPARENCY}

The simulations are completed with the 1D3V relativistic
electromagnetic Particle-In-Cell (PIC) code LPIC++~\cite{lpic}. In
the simulations, the initial electron temperature is set to be zero. There are 200 macroparticles per cell. Flat-shaped p-polarized laser irradiation of an initially homogeneous plasma foil with incidence angle  $24.62^{\circ}$ is considered.
All the simulation data is obtained in the boosted frame~\cite{bourdier:1804}, where oblique incidence in laboratory frame is transformed to normal incidence.

\begin{figure}

\includegraphics{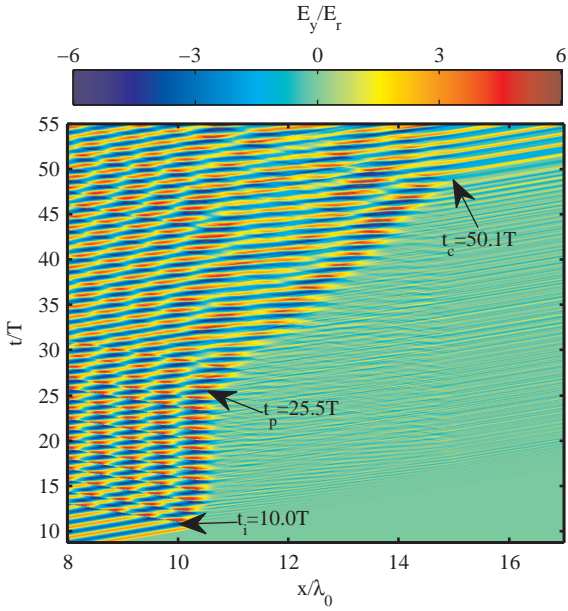}

\caption{\label{fig:p5_n5_ey} The spacetime
distribution of the laser field $E_y$. The laser starts to penetrate into the plasma
at $t_p=25.5T$ with a time interval $\Delta t=15.5T$ after the irradiating time of $t_i=10.0T$,
and the whole plasma becomes transparent at $t_c=50.1T$.}
\end{figure}

In the first simulation, a laser pulse with amplitude $a_0=E_0/E_r=5$
irradiates a thin plasma with initial electron density $n_e=5n_c$, where $E_r=m_e \omega c/e$ and $n_c=m_e\omega^{2}/(4\pi e^2)$. The width L of the
plasma is $5\lambda_0 $($\lambda_0$ is the wave length of the laser light in vacuum), and the plasma starts from $10.0\lambda_0$ to $15.0\lambda_0$ in the simulation box. The ions are fixed. According to the SIT theory, the laser amplitude is smaller than the SIT threshold $a_{lin}\approx 4n_e/(\pi n_c)=6.4$~\cite{guerin:2693}, the plasma should be opaque to the laser. In our simulation (Fig.~\ref{fig:p5_n5_ey}), the laser arrives at the front surface of the plasma at time $t_i=10.0T$, and is reflected for $t<t_p=25.5T$. That is to say, the plasma is opaque for $t<t_p$. However, for $t>t_p$, the laser begins to penetrate into the plasma, and the whole plasma becomes transparent for the laser at $t_c=50.1T$.

\begin{figure}
\includegraphics[scale=1]{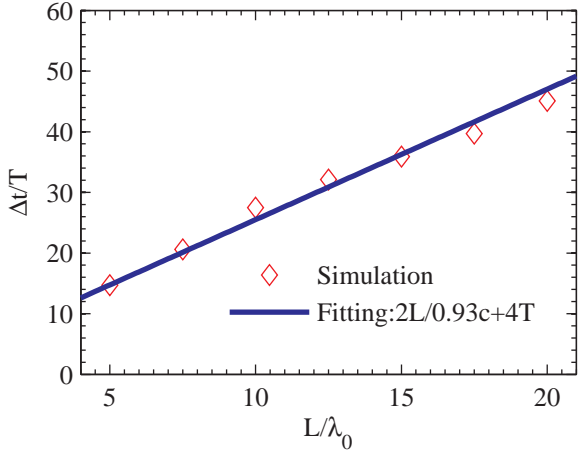}
\caption{\label{fig:L_T_nt} The time interval $\Delta t$ between
the laser irradiating time $t_i$ and penetrating time $t_p$ as a function of the plasma thickness L. }
\end{figure}

To explore the transition of overdense plasma from opacity to transparency, plasmas with different thicknesses are considered.
Simulation results indicate that the time interval defined as $\Delta t=t_p-t_i$ increases linearly with the increase of plasma thickness L, as shown in Fig.~\ref{fig:L_T_nt}. The dependence of $\Delta t$ on L is fitted as
\begin{equation}
 \Delta t=2L/0.93c+4T. \label{eq:1}
\end{equation}
Above equation indicates that $\Delta t$ is related to something which travels 2L distance at a velocity of 0.93c. In the next section, it will be demonstrated that this "something" is the hot-electron refluxing.

\section{Refluxing induced transparency}

\begin{figure}
\includegraphics[scale=1]{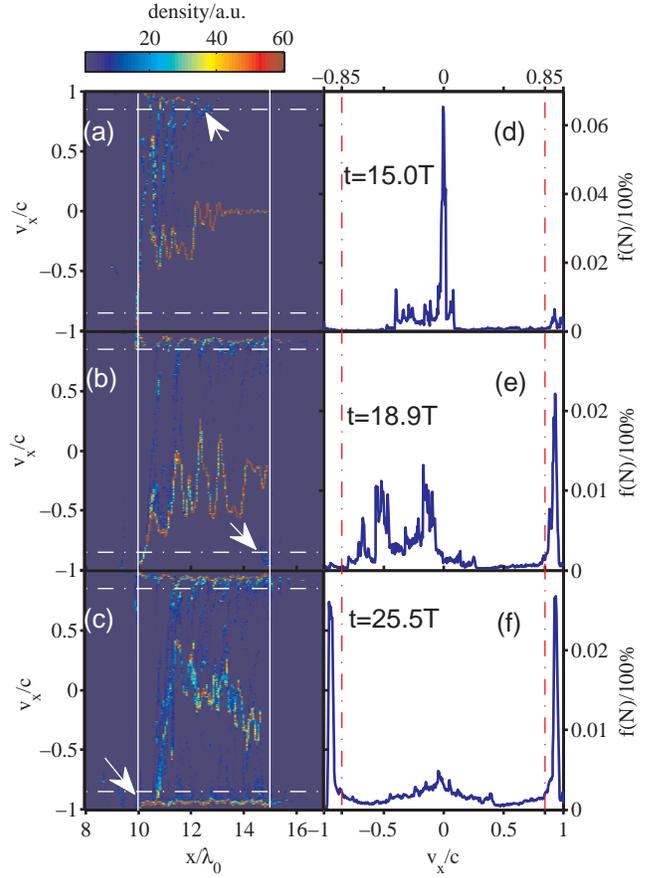}
\caption{\label{fig:phasespace} The phase-space distribution of the electrons (a)-(c)) and corresponding velocity distribution (d)-(f)
at $t=15.0T$, $t=18.9T$ and $t=25.5T$ respectively  with the same simulation parameters as Fig.~\ref{fig:p5_n5_ey}
(where the solid lines indicate the plasma's location, the dash-dot lines mark $|v_x|=0.85c$
and the three white arrows show the front position that the hot electrons arrive at three different time ).}
\end{figure}

To illustrate that it is the hot-electron refluxing which induces the initially opaque plasma become transparent, the kinetics of hot electrons is investigated.
With the incidence of the p-polarized laser, hot electrons are generated at the plasma surface($x=10.0\lambda_0$) by vacuum heating~\cite{PhysRevLett.59.52}, $J\times B$ heating~\cite{kruer:430} and so on.
The phase-space distribution and velocity distribution of the electrons at time $t=15.0T$, $t=18.9T$ and $t=25.5T$ are shown in Fig.~\ref{fig:phasespace}.

For $t=15.0T$, the phase-space distribution of the electrons indicates that hot electrons with longitudinal velocity $v_x>0.85c$ are generated and keep moving toward the rear surface($x=15.0\lambda_0$) of the plasma (Fig.~\ref{fig:phasespace}(a)). The corresponding velocity distribution of electrons shows a peak at $v_x>0.85c$ (Fig.~\ref{fig:phasespace}(d)). For $t=18.9T$, more hot electrons with $v_x>0.85c$ are generated and the front of the hot electrons reach the rear surface, as shown in Fig.~\ref{fig:phasespace}(b) and (e). According to Ref.~\onlinecite{myatt:056301}, $90\%$ of the hot electrons will be reflected by the sheath field and form the hot-electron refluxing. For $t=25.5T$, the hot-electron refluxing arrives at the front surface as marked by the arrow in Fig.~\ref{fig:phasespace}(c). And the velocity distribution of electrons Fig.~\ref{fig:phasespace}(f) has two peaks in $v_x>0.85c$ and $v_x<-0.85c$ respectively. Average velocity of hot electrons with velocity $|v_x|>0.85c$ is calculated to be 0.93c. Two important results can be obtained from these figures,
firstly, the hot-electron refluxing arrives the front surface at time $t=25.5T$, consistent with the time $t_p$ the laser begins to penetrate into the plasma in Fig.~\ref{fig:p5_n5_ey}; secondly, the average velocity of the hot electrons is 0.93c, consistent with the denominator in the first term on the right of Eq.~\ref{eq:1}.

\begin{figure}
\includegraphics[scale=1]{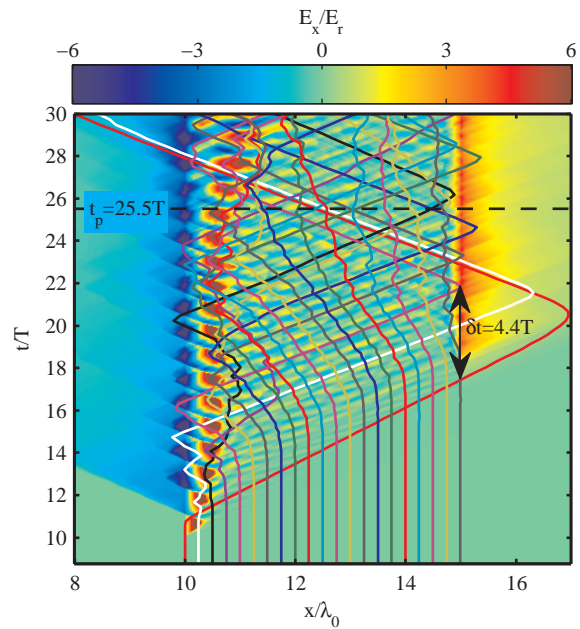}
\caption{\label{fig:refluxing}
The trajectories of 21 electrons (solid lines) and the spacetime distribution of electrostatic field $E_x$.
}
\end{figure}

As the hot-electron refluxing plays an important role in laser penetration, the generation and transporting of the refluxing deserves more attention.
Fig.~\ref{fig:refluxing} illustrates the trajectories of 21 electrons which are initially uniformly distributed in the plasma, and spacetime distribution of electrostatic field $E_x$ is shown in the same figure. When the laser irradiates the plasma at $t_i=10.0T$, the electrons on the plasma front surface are heated and move toward the rear surface. To balance the charge equilibrium, negative-directing return current is generated with velocity much smaller than the hot electrons~\cite{beg:2806}. As the hot electrons move out of the rear surface, a strong sheath electrostatic field is induced. The hot electrons are pulled back by the sheath electrostatic field and a hot-electron refluxing is formed in a time interval $\delta t=4.4T$ after the arrival of the hot-electron front at the rear surface. This picture of the generation of hot-electron refluxing is consistent with the results in Ref.~\onlinecite{PhysRevLett.74.2002} (see FIG. 1 there).

From above analysis, Eq.~\ref{eq:1} can be well understood. 0.93c is the average velocity of the hot electrons, 2L is the distance that the hot electrons travel on a round trip in the thin plasma, 4T is the average time the hot electrons lose at the rear surface.
These results confirm the assumption that it is the hot-electron refluxing which makes the overdense plasma transparent to the laser.

\section{REFLUXING ENHANCES TRANSPARENCY }

\begin{figure}
\includegraphics[scale=1]{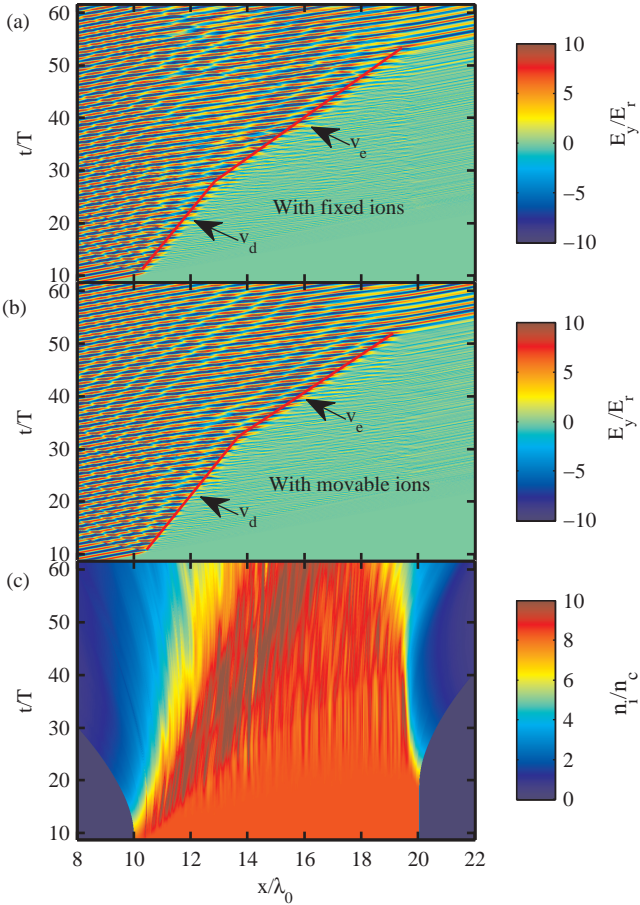}
\caption{\label{fig:Ey_ni} The laser has two different penetrating velocities
for direct penetration($v_d$) and refluxing enhanced penetration($v_e$) for $a_0=8,n_e=5n_c$  with fixed ions (a),
and movable ion namely with $m_i/m_e=3672$ (b).
(c) The spacetime distribution of the ions for the same parameters as (b).
}
\end{figure}

In the previous sections, the amplitude of the laser pulse is set to be $a_0=5$ and the ions are fixed. In this section, the results with higher laser intensity are investigated and the motion of ions is considered.

When the amplitude of the laser pulse is higher than the SIT threshold $a_{lin}$, the laser can
penetrate into the plasma directly with velocity $v_d$ as shown in Fig.~\ref{fig:Ey_ni}(a),
where $a_0=8$ and the ions are fixed. When the hot-electron refluxing returns to the laser penetration front,
it enhances the penetration and the laser gains a higher penetrating velocity $v_e$.
When the motion of ions is considered, same results is obtained, as shown in Fig.~\ref{fig:Ey_ni}(b).  Fig.~\ref{fig:Ey_ni}(c) shows the expansion of the ions.

\begin{figure}
\includegraphics[scale=1]{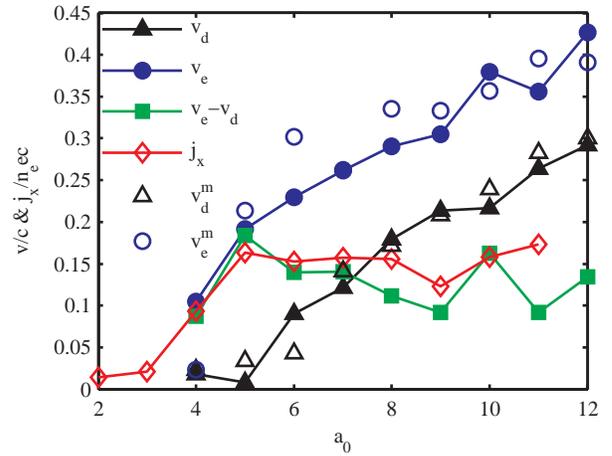}
\caption{\label{fig:vp_n5_refluxing_ion} The penetrating velocities for different laser intensities $a_0$ with and without the presence of hot-electron refluxing, and the current density $j_x$ of the hot-electron refluxing as a function of $a_0$.
Solid and open symbols of circle and triangle represent the simulation results for fixed ions and movable ions($m_i/m_e=3672$) respectively.
}
\end{figure}

The laser penetrating velocities for different $a_0$ with and without the presence of hot-electron refluxing have been summarized in Fig.~\ref{fig:vp_n5_refluxing_ion}.
The penetration velocity enhanced by the hot-electron refluxing has an increment $v_e-v_d$.
The hot-electron refluxing current density $j_x/n_eec$  has also been plotted as a function of laser intensities $a_0$. It is noteworthy that $v_e-v_d$ seems have the same trend with $j_x$, which means that the hot-electron refluxing has direct impact on the transparency of the plasma.
When ion motion is considered, for different $a_0$, the direct penetration velocity $v_d^m$ and the hot-electron refluxing enhanced $v_e^m$ consist well with $v_d$ and $v_e$ respectively.
These results indicate that the hot-electron refluxing can enhance laser's propagation in an overdense plasma.

\begin{figure}
\includegraphics[scale=1]{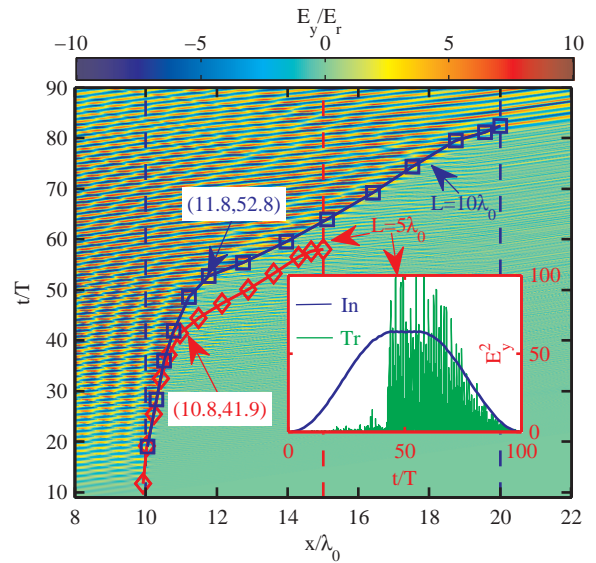}
\caption{\label{fig:Ey_Tr_In} The spacetime distribution of the laser field $E_y$ for plasma thickness $L=10\lambda_0$
and the trajectories of laser front for $L=10\lambda_0$ and $L=5\lambda_0$ respectively.
The incident laser pulse intensity envelope(In) and transmitted pulse(Tr) for $L=5\lambda_0$  are shown in the inset.}
\end{figure}

For a practical application, we consider a laser pulse with sinusoidally rising and falling profile (In), as shown in the inset of Fig.~\ref{fig:Ey_Tr_In}.
The results with different plasma thickness ($L=5\lambda_0$ and $L=10\lambda_0$) are shown in Fig.~\ref{fig:Ey_Tr_In}. For $L=5\lambda_0$, the penetration of the laser begins to speed up at time $t_1=41.9T$. While for $L=10\lambda_0$, this begins at time $t_2=52.8T$. The time interval between $t_1$ and $t_2$ is $\Delta t_1=t_1-t_2=10.9T$. $\Delta t_1$ is consistent with the estimation calculated according to Eq.~\ref{eq:1}, i.e., $\Delta t_1=2\Delta L/0.93c=10.8T$, where  $\Delta L=5\lambda_0$ is the thickness difference between the two plasmas. As the inset shown, the transmitted pulse (Tr) has a sharp front, which may be potentially useful for laser shaping~\cite{PhysRevLett.107.265002}. By measuring the reflected and transmitted laser intensities in experiments~\cite{2012NatPh...8..763P}, the hot-electron refluxing enhanced transparency may be verified by experiments.

\section{Conclusions}

In conclusion, using PIC simulations we have shown that the hot-electron refluxing can enhance relativistic transparency of overdense plasmas significantly.
When a p-polarized laser with laser intensity lower than the SIT threshold obliquely irradiates a thin overdense plasma, the laser cannot propagate through the plasma directly. But the arrival of the hot-electron refluxing can induce the laser's penetration. When the laser intensity is above the SIT threshold, the laser can penetrate into the plasma directly,  and the arrival of the hot-electron refluxing enhances the penetration velocity of the laser. These results are verified by the simulations with fixed ions and movable ions($m_i/m_e=3672$) respectively. The hot-electron refluxing enhanced transparency is also confirmed for a laser pulse with sinusoidally rising and falling profile for a practical application, and the transmitted laser pulse has a sharp front, which may be potentially useful for laser shaping.


%

\end{document}